\documentstyle[aps,prl,preprint,epsfig,amssymb]{revtex}
\begin{document}
\draft
\title{Finite-size scaling in the canonical ensemble}
\author{Youjin Deng~$^{1}$ and Henk W.J. Bl\"ote~$^{1,2}$}
\address{$^{1}$Faculty of Applied Sciences, Delft University of
Technology, P. O. Box 5046, 2600 GA Delft, The Netherlands}
\address{$^{2}$ Lorentz Institute, Leiden University,
  P. O. Box 9506, 2300 RA Leiden, The Netherlands}             
\date{\today} 
\maketitle 
\begin{abstract}
We investigate the critical scaling behavior of finite systems in the
canonical ensemble. The essential difference with the grand canonical
ensemble. i.e., the constraint on the number of particles, is already
known to lead to the Fisher renormalization phenomenon that modifies
the thermal critical singularities. We show that, in observables that
are not Fisher renormalized, it also leads to a finite-size effect
governed by an exponent $y_1$ that depends on the temperature exponent
$y_t$ and the dimensionality $d$ as $y_1=-|2y_t-d|$.
We verify this prediction by a Monte Carlo analysis of several 
two-dimensional lattice models in the percolation, the Ising and the
3-state Potts universality classes.
\end{abstract}

The central task of statistical physics is to calculate the partition
function and the thermodynamic observables. Different ensembles
may be employed to this purpose, which then naturally yield different
forms of the thermodynamic functions. In the thermodynamic limit, the
different approaches generally yield equivalent results for the relations
between the  thermodynamic variables. However, differences occur in
finite systems. We illustrate this by means of a lattice gas,
described by a finite size $L$, a temperature $T$ and a third parameter
governing the number of particles.  In the canonical ensemble this 
third parameter is the particle density $\rho$.
A calculation of an observable $A$ in the canonical ensemble will thus,
in principle, yield its expectation value $\langle A \rangle_c$ as a
function $A^{(c)}(T,\rho,L)$.
In the grand ensemble, one employs the chemical potential $\mu$ as the
third parameter. One may thus calculate the expectation value
$\langle A \rangle_g = A^{(g)}(T,\mu,L)$ as well as the grand canonical
density $\langle \rho \rangle_g = \rho^{(g)}(T,\mu,L)$.
In the thermodynamic limit one expects in general that
\begin{equation}
 A^{(g)}(T,\mu,L) = A^{(c)}(T,\rho^{(g)}(T,\mu,L),L) 
\mbox{\hspace{5mm}} (L \to \infty) ,
\label{Tlim}
\end{equation}
for expectation values of the form
$\langle A \rangle=\sum_\Gamma A(\Gamma) P(\Gamma)$, where 
$P(\Gamma)$ is the probability of state $\Gamma$ in the pertinent ensemble.
It does {\em not} apply to quantities obtained by differentiation of
observables to  $T$, such as the specific heat. This is evident when
we define $A' \equiv \partial A / \partial T$:
\begin{equation}
 {A'}^{(g)}(T,\mu,L) = {A'}^{(c)}(T,\rho^{(g)}(T,\mu,L),L)
+\frac{\partial A}{\partial \rho} \frac{\partial \rho}{\partial T} .
\label{Apri}
\end{equation}
The last term violates Eq.~(\ref{Tlim}). It leads to the Fisher
renormalization \cite{MEF} effect that may affect even the exponents
of leading critical singularities. 

In this paper, we point out that, in the canonical ensemble, new
finite-size effects appear also for quantities that do not involve
differentiations to $T$ or other temperature-like variables.
They appear because substituting $\rho^{(g)}$ in the right-hand
side of Eq.~(\ref{Tlim}) is not precisely equivalent with taking the
grand canonical expectation value.
Namely, in the grand ensemble, $\rho$ is still allowed to fluctuate,
which is not the case in the canonical ensemble.
We derive this new finite-size effect for a $d$-dimensional lattice
gas, described by a reduced Hamiltonian ${\mathcal H}$ with
variables $\sigma_i=0$ (1) denoting the absence (presence) of
a particle on lattice site $i$.
The grand partition sum is
\begin{equation}
Z^{(\rm g)}(T, \mu,L) = \sum_{\sigma_1=0}^{1} \sum_{\sigma_2=0}^{1} \cdots
\sum_{\sigma_N=0}^{1} \exp[-{\mathcal H}]  ,
\label{ZG}
\end{equation}
where the sum is performed independently on all $N\equiv L^d$ lattice gas 
variables. The particle density $\rho$ follows from differentiation
of Eq.~(\ref{ZG}):
\begin{equation}
\rho(T,\mu,L) = \frac{1}{N} \frac{\partial \ln Z^{(\rm g)}}{\partial \mu} .
\label{rho}
\end{equation}
Other thermodynamic quantities $A$ can be obtained similarly by
differentiation to the conjugate parameter. This leads to the following
form for the expectation value of an observable in the grand ensemble 
\begin{equation}
A^{(\rm g)}(\mu,T) =
 \sum_{\{\sigma\}} A(\{\sigma\})\exp[-{\mathcal H}]/Z^{(\rm g)} .
\label{AG}
\end{equation}
The sum on $\{\sigma\}$ is shorthand for the sums in Eq.~(\ref{ZG}).

The canonical partition sum is
\begin{equation}
Z^{(\rm c)}(\rho,T) = \sum_{\{\sigma\}} \delta_{N_p,\sum_k \sigma_k}
\exp[-{\mathcal H}] ,
\label{ZC}
\end{equation}
where $N_p \equiv N \rho$ is the particle number and
the Kronecker $\delta$ imposes the constraint. 
The canonical expectation value of $A$ is 
\begin{equation}
A^{(\rm c)}(\rho,T) = \sum_{\{\sigma\}} \delta_{N\rho,\sum_k \sigma_k}
A(\{\sigma\}) \exp[-{\mathcal H}] / Z^{(\rm c)} .
\label{AC}
\end{equation}
Therefore, $A^{(\rm g)}$ and $A^{(\rm c)}$ are related as
\begin{eqnarray}
A^{(\rm g)}(\mu,T) &=& \frac{\sum_{N_p=0}^{N} 
\sum_{\{\sigma\}} \delta_{N_p,\sum_k \sigma_k} A(\{\sigma\})
\exp[-{\mathcal H}]}{Z^{(\rm c)}} \, \frac{Z^{(\rm c)}}{Z^{(\rm g)}} 
\nonumber \\
 &=& \sum_{N_p=0}^{N} A^{(\rm c)}(N_p/N, T) P(\mu,N_p/N)  ,
\label{AGC}
\end{eqnarray}
where the grand probability $P(\mu,\rho)$ that a particle
density $\rho$ occurs is equal to the ratio $Z^{(\rm c)}/Z^{(\rm g)}$.
In the thermodynamic limit, Eqs.~(\ref{AG}) and (\ref{AC}) should
be equivalent as long as $\mu$ and $\rho$ satisfy Eq.~(\ref{rho}).

For simplicity, let $A$ be a quantity whose leading singular
finite-size-scaling term is a constant, such as a the Binder
ratio \cite{Binder} or 
other dimensionless finite-size amplitude ratios.
Near a critical point $\rho=\rho_c$, $T=T_c$, 
its finite-size scaling behavior reads \cite{MEF,YD3}
\begin{equation}
A^{(\rm c)}(\rho,T_c,L) = A_c^{(\rm c)} +
\sum_k a_k (\rho-\rho_c)^k L^{ky_{\rho}} +b L^{y_1}  ,
\label{ACs}
\end{equation}
where the last term allows for a finite size effect at $T_c$.
Eq.~(\ref{AGC}) thus becomes 
\begin{equation}
A^{(\rm g)}(\mu,T_c,L) = A_c^{(\rm c)} +
\sum_k a_k\langle (\rho-\rho_c)^k\rangle L^{ky_{\rho}} +b L^{y_1} ,
\label{AGs0}
\end{equation}
where $\langle(\rho-\rho_c)^n\rangle= \sum_{N_p=0}^{N}(\rho-\rho_c)^n
P(\mu,N_p/N)$ is the $n$-th moment of $(\rho-\rho_c)$ in the grand
ensemble.  The second moment scales as
\begin{equation}
\langle(\rho-\rho_c)^2\rangle = r_0 L^{-d} + r_1 L^{2y_t-2d} .
\label{rhos}
\end{equation}
At criticality, the term  with $k=1$ in Eq.~(\ref{AGs0}) is
suppressed, and substitution of Eq.~(\ref{rhos})  yields
\begin{equation}
A^{(\rm g)}(\mu_c,T_c,L) = A_c^{(\rm c)} +w L^{-|2y_t-d|} +b L^{y_1} 
+\cdots 
\label{AGs2}
\end{equation}
where we have used $y_{\rho}=d-y_t$ for $2y_t-d \geq 0$ and 
$y_{\rho}=y_t$ for $2y_t-d <0$ 
according to the Fisher renormalization~\cite{MEF,YD3}.
The asymptotic value 
is $A_c^{(\rm g)}= A_c^{(\rm c)}$ for $2y_t -d<0$ and
$A_c^{(\rm g)}= A_c^{(\rm c)}+a_2 r_1$ for $2y_t -d>0$. 
Eq.~(\ref{AGs2}) should match the finite-size scaling formula of $A$
in the grand canonical ensemble. Here we are on the firm ground of
the scaling properties of the free energy,
in terms of scaling fields associated with the intensive
thermodynamic parameters.
Differentiation to appropriate fields thus yields the scaling
behavior~\cite{FSS} as 
\begin{equation}
A^{(\rm g)}(\mu_c,T_c,L) = A_c^{\rm (g)} + gL^{y_i} +\cdots .
\end{equation}
Since the irrelevant exponent $y_i$ is in general not equal to 
$-|2y_t-d|$, comparison with Eq.~(\ref{AGs2}) shows that the last two
terms in Eq.~(\ref{AGs2}) must cancel one another, i.e.,  
\begin{equation}
y_1=-|2y_t-d| .
\label{y1}
\end{equation}
This accounts for the new  correction exponent in Eq.~(\ref{ACs}). 
Further, if the asymptotic value $A_c^{(\rm g)}$ is universal, one
expects that this universal value applies to the canonical
ensemble as well if $2y_t -d<0$.

We test these predictions numerically, by means of Monte Carlo
simulations of five models with periodic boundary conditions,
belonging to the $d=2$,
$q=1$, 2 and 3 Potts universality classes. These models are:

\noindent
(1) the hard-square lattice gas model 
\begin{equation}
{\cal H} = -K \sum_{\langle nn \rangle } \sigma_i \sigma_j
- \mu \sum_k \sigma_k \hspace{5mm} (\sigma=1,0) ,
\label{hardhh}
\end{equation}
where the sum is on all nearest-neighbor pairs. 
The particles $\sigma=1$ have a `hard' core so that
nearest-neighbor exclusion applies,
i.e., $K \rightarrow - \infty$.
The critical chemical potential is~\cite{WAG} 
$\mu_c=1.334 \; 015 \; 100  \;277 \; 74(1)$ with particle
density $\rho_c=0.367 \; 742 \; 999 \; 041 \; 0 (3)$.

\noindent
(2) the Blume-Capel model~\cite{MB} on the square lattice
\begin{equation}
{\cal H} = -K \sum_{\langle nn \rangle } s_i s_j
- \mu \sum_k s_k^2  \hspace{5mm} (s=0, \pm 1) .
\label{HamBC}
\end{equation}
Zero spins $s=0$ are called `vacancies'.
For $K=1$ the critical point~\cite{XFQ} lies at
$\mu_c (K=1)=-1.702\, 717\, 8 (2)$, with a vacancy density
$\rho_{vc}=0.349\,583\, 0(2)$.

\noindent
(3) the dilute 3-state Potts model on the square lattice
\begin{equation}
{\cal H} = -K \sum_{\langle nn \rangle }
\delta_{\sigma_i, \sigma_j } (1-\delta_{\sigma_i, 0})
+ \mu \sum_i \delta_{\sigma_i, 0}  .
\label{Hampot}
\end{equation}
Each site carries a Potts variable $\sigma=1,\cdots,q$,
or a vacancy $\sigma=0$.  The chemical potential $\mu$ controls
the number of vacancies $\langle N_v \rangle $.
A critical point was located~\cite{XFQ} at $K=1.16940$,
$\mu_c=-1.376 \, 413(4)$, with a 
vacancy density $\rho_{vc}=0.105\, 273\, 0(2)$. 

\noindent
(4) Baxter's hard-hexagon model \cite{RJB0}, also described by
Eq.~(\ref{hardhh}) with $K \rightarrow - \infty$, but defined 
on the triangular lattice.
Its critical point is \cite{RJB0} 
$\mu_c= \ln [(11+5\sqrt{5})/2]$, with particle
density $\rho_c= (5-\sqrt{5})/10$. 

\noindent
(5) the bond percolation model on the square lattice, whose
percolation threshold occurs at bond probability $p_c=1/2$.

For these models we sampled dimensionless quantities $Q$ defined on
the probability distribution of the order parameter, which is the
magnetization $m$ for Isinglike models such as the Blume-Capel model.
For that case we employ the Binder ratio \cite{Binder}
\begin{equation}
Q= \langle m^2 \rangle^2/\langle m^4 \rangle .
\label{QBC}
\end{equation}
For the hard-square model, the same definition of $Q$ applies
with $m$ replaced by the difference $\rho_1-\rho_2$ of the 
particle densities on the two sublattices.  For the dilute 3-state
Potts model and the hard-hexagon model we used Eq.~(\ref{QBC}) with
$m^2$ replaced by
$\rho_1^2+\rho_2^2+\rho_3^2-\rho_1\rho_2-\rho_2\rho_3-\rho_3\rho_1$
where $\rho_i$ denotes the density of the Potts variables in state $i$,
and the particle density on sublattice $i$ respectively. The mapping
of the Ising model
on the random-cluster model \cite{PWK} makes it possible to express
$Q$ in moments of the cluster size distribution as
\begin{equation}
Q= \langle l_2 \rangle^2/[3\langle l_2^2 \rangle-2
\langle l_4 \rangle]
\label{Qrc}
\end{equation}
with the moments of the cluster size distribution defined as
$l_k = L^{-2k}  \sum_{i=1}^{N_{cl}} c_i ^k$
where $c_i$ is the size of the $i$th cluster, and $N_{cl}$ is the
total number of clusters. Eq.~(\ref{Qrc}) defines
$Q$ for the bond percolation model.

The conservation of particles in the canonical ensemble requires
a definition of a `particle'. This meaning is obvious for
both lattice gases. In the percolation case,
occupied bonds serve as `particles': our simulations conserve the
number of occupied bonds as $N_b=L^2$. This is simply realized by
Monte Carlo moves that randomly interchange an occupied bond and
an empty one. For the Blume-Capel and the Potts model
we interpret nonzero spins $s_i\neq0$ and variables $\sigma_i\neq0$
as particles. All simulations, except for the bond percolation model,
used the geometric Monte Carlo algorithm \cite{JRH0} that is well
suited because it conserves the particle number and suppresses critical
slowing down.  For the Blume-Capel 
and the dilute Potts model it was combined with Wolff~\cite{UW}
cluster steps that do not affect the vacancies.

For the bond percolation model, $L^2/4$ pairs of edges were interchanged 
between two subsequent samples. The sampling procedure decomposed the
whole lattice into percolating clusters. 
The system sizes took $20$ values in the
range $ 4 \leq L \leq 4000$. About $4 \times 10^7$ samples
were taken for each $L$ for $L \leq 800$, and $4 \times 10^6$ samples
for each $L$ for $L >800$. 
For the hard-square lattice gas,
we simulated systems up to $L=4000$, and for the hard-hexagon gas
up to $L=1680$.
We took about $2 \times 10^7$ or more samples for each system size.
For the Blume-Capel model and the $q=3$ dilute Potts model, 
we used systems up to $L=1600$ and $L=800$, respectively; 
about $2 \times 10^7$ or more samples were taken for each system size.
Since in a finite system the critical number of the particles
$\rho_c L^2$ is not an integer, we obtained the data at criticality
by interpolation. For each of these models we determined $Q$ 
at criticality, using only modest computer resources.

\begin{figure}
\begin{center}
\leavevmode
\epsfxsize 7.5cm
\epsfbox{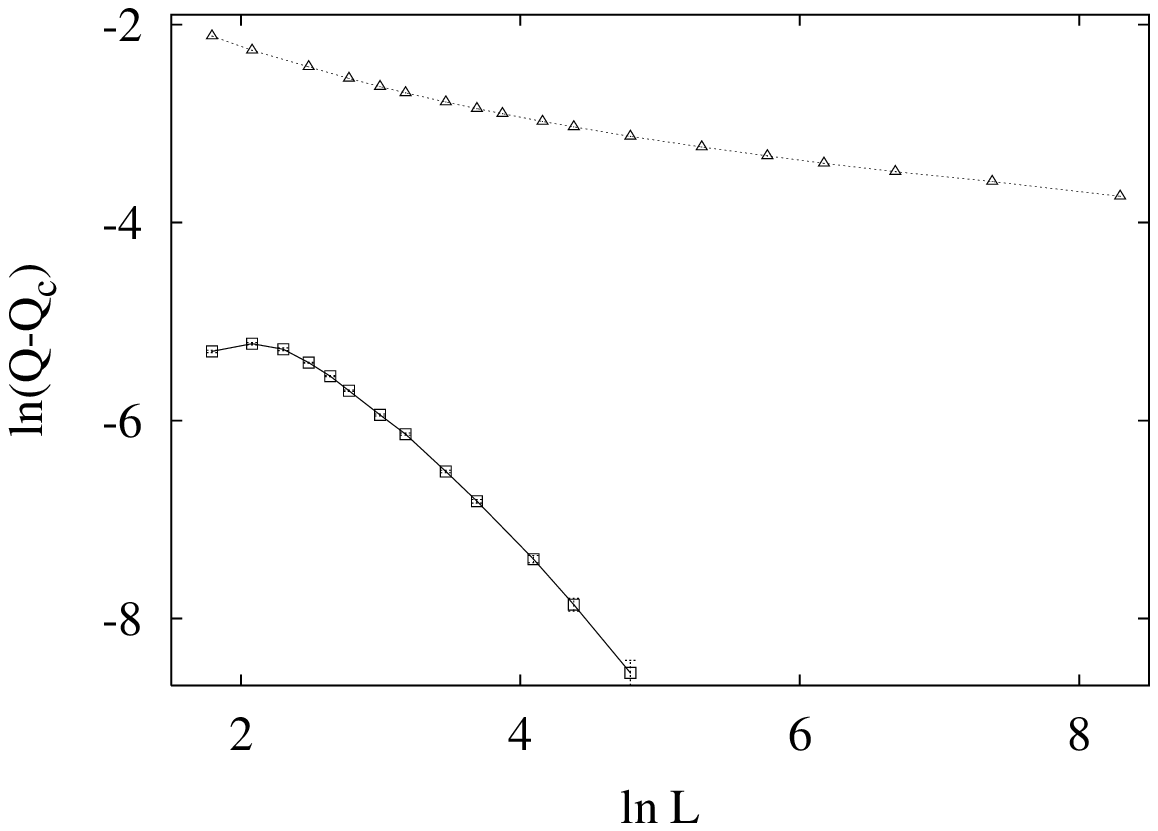}
\end{center}
\caption{Finite-size dependence of $Q$ of the canonical $(\triangle)$
and the grand canonical $(\Box)$ hard-square lattice gas. }
\label{fig01}
\end{figure}

\begin{figure}
\begin{center}
\leavevmode
\epsfxsize 7.5cm
\epsfbox{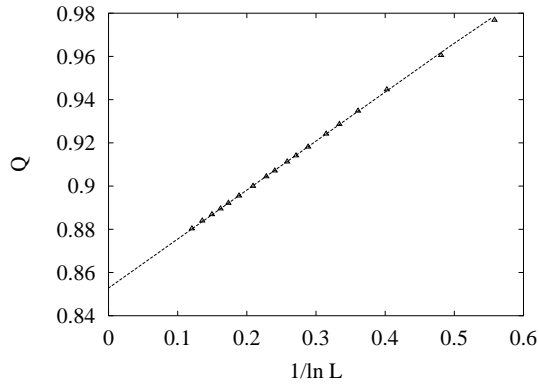}
\end{center}
\caption{Ratio $Q$ of the canonical hard-square gas
vs. $1/\ln L$.} 
\label{fig02}
\end{figure}

Fig.~\ref{fig01} compares the finite-size corrections in $Q$ of
the canonical and the grand canonical hard-square lattice gas. 
The corrections are much more prominent in the canonical
case, and decay more slowly with $L$.
In the grand ensemble, $Q$ scales as 
$Q=Q_c+ b_i L^{y_i} + b_a L^{y_a}$ with an irrelevant Ising exponent
$y_i=-2$ and an exponent $y_a=2-2y_h=-7/4$ describing the analytic
background in $\langle m^2\rangle$. 
The canonical $Q(L)$ data are shown versus $1/\log L$ in Fig.~\ref{fig02};
the approximate linearity
indicates that the leading finite-size dependence is of a logarithmic 
nature. This is in line with Eq.~(\ref{y1}), which yields 
$y_1=0$ for Isinglike models.  We thus fitted the  $Q$ data by
\begin{equation}
Q=Q_c+ \sum_{k=1}^2 b_k(\ln L-\ln L_0)^{-k} +b_i L^{y_i} +b_a L^{y_a},
\label{fitq2}
\end{equation}
where $b_k$ and $L_0$ are constants, with fixed exponents 
$y_i=-2$ and $y_a=2-2y_h =-7/4$.
The $Q$ data in the range $8 \leq L \leq 4000$ are indeed well
described by Eq.~(\ref{fitq2}).  We obtain $Q_c=0.8563(2)$, $L_0=3.0(1)$,
$b_1=0.190 (2)$, and $b_2=-0.102(6)$; the value of $Q_c$ is in agreement
with the grand canonical value $Q_c=0.856216(1)$~\cite{GK,JS}.
We also fitted without logarithmic corrections, by 
\begin{equation}
Q(L)=Q_c+b_1 L^{y_1} +b_a L^{y_a} +b_2 L^{y_i} +b_3 L^{y_i-1} ,
\label{fitq1}
\end{equation}
which was unsatisfactory because
only a narrow range of sizes $L$ could be accommodated, while
the estimate $y_1=-0.15(3)$ of the exponent is close to 0.   
Also for the Blume-Capel model we observed slowly decaying finite-size 
dependence in $Q$.  The $Q$ data for $12 \leq L \leq 1600$ were 
satisfactorily fitted by Eq.~(\ref{fitq2}). We obtain 
$Q_c=0.8567 (7)$ and $L_0=3.3(1)$.  

The $Q$ data for
the hard-hexagon model were fitted by Eq.~(\ref{fitq1})
with exponents 
$y_a=2-2y_h =-26/15$ and $y_i=-4/5$~\cite{BN2}. The fits yield
$Q_c=0.9223(5)$ and  $y_1=-0.403 (6)$ for the lattice gas. 
We have also determined the grand canonical value $Q_c=0.85670 (4)$, 
clearly different from the canonical value. 
For a similar fit to the data for the $q=3$ dilute Potts model we find
$Q_c=0.9190(5)$ and $y_1=-0.408 (10)$.

The $Q$ data for the bond-percolation model were fitted by
Eq.~(\ref{fitq1}) with exponents 
$y_a=2-2y_h=-43/24$ and $y_i=-2$~\cite{BN2}. We obtain 
$Q_c=0.87052(8)$ and $y_1=-0.499 (2) \approx -1/2$. 
For the grand canonical version of this model,
we find~\cite{YD4} that the $Q(L)$ data are also well
described by Eq.~(\ref{fitq1}), but without the term with 
exponent $y_1=-1/2$.  The universal ratio 
$Q_c=0.87053(2)$~\cite{YD4} agrees well with
the canonical value.

Our numerical analyses of $Q$, together with the predictions of 
Eq.~(\ref{y1}),  are summarized in Table \ref{tab_1}.

\begin{figure}
\begin{center}
\leavevmode
\epsfxsize 7.5cm
\epsfbox{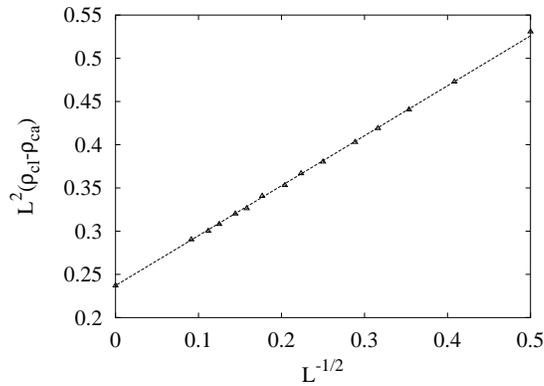}
\end{center}
\caption{Scaled finite-size dependence of the cluster-number density
$\rho_{cl}$ of the canonical bond-percolation model vs. $L^{-1/2}$. }
\label{fig03}
\end{figure}

Finite-size effects with exponent $y_1=-|2y_t-d|$ are not
restricted to dimensionless quantities. Our derivation applies 
generally to the leading singular finite-size-dependent terms,
which are modified by a similar factor as in the case of $Q$.
We tested this numerically for the susceptibility (or its analog)
of each of the the five models. Indeed we find effects in the
critical canonical susceptibilities $\chi^{(c)}$ according to 
$\chi^{(c)} (L)=\chi^{(g)}(L) (1+ r L^{y_1}+\cdots) $.
A further verification concerns
the cluster-number density $\rho_{cl}=N_{cl}/N$ of the
bond-percolation model,  which 
at criticality behaves as~\cite{RMZ0} 
$\rho_{cl} (L)=\rho_{ca}+L^{-2} (b_0 +b_1 L^{-2} +\cdots) $
in the grand canonical ensemble, with
$\rho_{ca}=(3\sqrt{3}-5)/2$  \cite{RMZ0}. 
The $\rho_{cl} (L)$ data in the canonical ensemble, shown in
Fig.~\ref{fig03}, agree with $y_1=-1/2$.
The fit of the canonical data by
\begin{equation}
\rho_{cl} (L)=\rho_{ca}+ L^{-2} (b_0 +b_1 L^{y_1}+b_2 L^{-2}), 
\label{fitnc1}
\end{equation}
confirms this value as $y_1=-0.51(1) \approx -1/2$.

As another generalization, we expect finite-size effects with
exponent $-|2y_h-d|$ in systems with a conserved magnetization.

The authors are indebted to Michael E. Fisher, Jouke R. Heringa, 
and Xiaofeng Qian for valuable discussions.
This research was supported by the Dutch FOM foundation (``Stichting
voor Fundamenteel Onderzoek der Materie'') which is financially supported by
the NWO (`Nederlandse Organisatie voor Wetenschappelijk Onderzoek')

\begin{table}
\caption{Summary of the main numerical results. The superscripts 
(n) and (t) indicate the numerical estimate and the theoretical 
prediction, respectively.  }
\label{tab_1}
\begin{center}
\begin{tabular}{||l|l|l|l|l||}
model            &$y_1^{\rm (n)}$ &$y_1^{\rm (t)}$&$Q_c^{(c)}$&$Q_c^{(g)}$  \\
\hline
hard square        & $\approx 0$  &     0     &  0.8563 (2)  & 0.856216 (1) \\
Blume-Capel        & $\approx 0$  &     0     &  0.8567 (8)  & 0.856216 (1) \\
$q=3$ Potts        & $-0.408$ (10)&  -2/5     &  0.9190 (5)  & --           \\
hard hexagon       & $-0.403$ (6) &  -2/5     &  0.9223 (5)  & 0.85670 (4) \\
bond percolation   & $-0.499$ (2) &  -1/2     &  0.8705 (1)  & 0.87053 (2) \\
\end{tabular}
\end{center}
\end{table}
\end{document}